**TITLE:**

Explicit pixel-value correction of lateral response artifacts in radiochromic film dosimetry: Evaluation of single-, double-, and triple-channel methods

**AUTHORS:**

Lixiang Guo[1,2*] and Ken Kang-Hsin Wang[1]

[1]Department of Radiation Oncology, University of Texas Southwestern Medical Center, Dallas, TX 75390, USA

[2]Department of Radiation Oncology, University of Florida, Gainesville, FL 32610, USA

***CORRESPONDENCE:**

Lixiang Guo

Department of Radiation Oncology, University of Florida, 2000 SW Archer Road, Gainesville, FL 32610, USA

Tel: 469-236-9619

Email: l.guo@ufl.edu

**RUNNING TITLE:**

Lateral response artifact correction for film dosimetry

**AUTHOR CONTRIBUTION STATEMENT**

**L.G.**: conceptualization, investigation, formal analysis, writing – original draft, writing – review & editing; **K.W.**: conceptualization, writing – review & editing.

**CONFLICTS OF INTEREST STATEMENT**

The authors have no relevant conflicts of interest to disclose.

**ACKNOWLEDGEMENTS**

The authors acknowledge funding support from the Cancer Prevention and Research Institute of Texas (RR200042) and National Institutes of Health (NCI R01 CA307552).

**ABSTRACT**

**Background:** Flatbed scanners introduce a lateral response artifact (LRA) along the axis perpendicular to lamp motion. Because the artifact varies with scanner position, pixel value (PV), color channel, and film model, it may violate the channel-common dose-independent RGB perturbation assumption of multichannel dosimetry and may not be removed by multichannel dosimetry alone.

**Purpose:** To characterize the PV-domain LRA of EBT4 and EBT-XD films, compare the Lewis two-film and Full film-cyclic LRA corrections, and evaluate their effects on single-, double-, and triple-channel dosimetry.

**Methods:** Seven films spanning the response range were scanned at seven lateral positions in a cyclic-shift design. For each position and RGB channel, the center-equivalent response was modeled as $PV_c = A + B \times PV$. The Lewis LRA correction used two films and linearly interpolated A and B across positions; the Full LRA correction used all center-local pairs and applied piecewise cubic Hermite interpolation (PCHIP) directly to $PV_c - PV$. Single-, double-, and triple-channel dosimetry methods were evaluated based on profile consistency between portrait scans (affected by LRA) and landscape scans (not affected by LRA), as well as on dose accuracy.

**Results:** The LRA was affine in PV but asymmetric about the scanner center and dependent on film model and channel. Both corrections reduced portrait-landscape profile differences. After correction, the mean absolute profile difference averaged across MU levels ranged from 1.4% to 5.3% for EBT4 and from 2.4% to 6.8% for EBT-XD for the different dosimetry methods investigated. The corrections performed similarly for EBT4; Full LRA correction produced lower profile discrepancies for several EBT-XD double- and triple-channel methods. Residual blue-channel discrepancies limited blue, red-blue, and triple-channel performance. At 500 MU and above, red channel showed the closest dose agreement with reference output; at 100 MU, both Lewis and Full corrections increased dose deviations for several single-channel estimates. At moderate-to-high MU, triple-channel deviations were generally smaller than blue and red-blue deviations. Full LRA also produced smaller deviations than Lewis LRA in several EBT-XD conditions.

**Conclusions:** The LRA should be corrected in the raw-PV domain before dose reconstruction. Compared with Lewis LRA correction, the Full correction provided better overall preservation of dose accuracy and profile consistency, particularly for EBT-XD. Improvements in profile consistency did not always improve central-dose agreement, so both outcomes should be commissioned and evaluated separately for LRA correction.

## 1. INTRODUCTION

Radiochromic film is widely used for two-dimensional dose verification because it offers high spatial resolution, near-tissue equivalence, minimal energy dependence across clinically relevant beam qualities, and negligible perturbation of the radiation field[1,2]. Its dosimetric performance, however, depends on the irradiation and readout workflow, including film handling, postirradiation growth, scanner response, calibration, and image analysis[1]. To achieve the high level of dosimetric accuracy required in radiation therapy (RT), the film and scanner should therefore be commissioned and maintained as an integrated system.

A major scanner-related limitation is the lateral response artifact (LRA). For a film exposed to a spatially uniform dose, the measured response varies with distance from the scanner midline along the axis perpendicular to the scanner lamp motion[3-5]. The artifact has been attributed to increasing optical-path obliquity as well as polarization- and scattering-dependent interactions between the scanner optics and the radiation-induced polymer structure of the film[3]. Its magnitude depends on scanner design, lateral position, color channel, film model, orientation, and darkening[3-5]. Consequently, a large film may exhibit systematic distortion of the measured profile from the scanner center toward the lateral edges.

Several strategies have been proposed to mitigate LRA. Lateral position specific calibration and response-rescaling methods explicitly map an off-center response to the corresponding response at the scanner center[4,5,10]. Lewis and Chan proposed a linear relationship between the pixel value (PV) measured at a lateral position and that measured at the scanner center, $PV_c$: $PV_c = A + B \times PV$, where A and B are fitting parameters. They further proposed linearly interpolating A and B between discrete calibration positions to obtain a continuous spatial correction[4]. More recent approaches reduce the calibration or readout burden through image stitching or accelerated acquisition protocols[12,13]. All these methods share a central principle: the spatially varying scanner-film response should be corrected before pixel value is converted to dose. The correction for LRA becomes important for large fields, where measurements extend far from the scanner midline.

Multichannel film dosimetry was developed to separate dose-dependent film response from dose-independent disturbances by exploiting the different sensitivities of the red, green, and blue channels[6-9]. These dose-independent disturbances arise from variations in film thickness, noise, dust, and other spatial nonuniformities, which are believed to be minimized by multichannel dosimetry. Conventional multichannel dosimetry formulations, however, assume that the dose-independent disturbances to RGB channels have a common structure and remain sufficiently small[8,9]. LRA does not necessarily satisfy these assumptions because its magnitude and PV dependence can differ among the channels. Multichannel processing may therefore attenuate LRA under selected conditions, but it cannot be assumed to completely eliminate the artifact, particularly at large lateral offset.

The introduction of EBT4 and the established use of EBT-XD for higher-dose applications motivate a direct comparison of their scanner-induced lateral response artifacts. EBT4 has reduced orientation dependence relative to earlier EBT generations, whereas EBT-XD has lower dose sensitivity and a wider applicable dose range[11,14-16]. Their distinct spectral and dose-response characteristics can cause the same scanner-induced LRA to produce different dose errors. A practical question is how an LRA correction characterized at discrete scanner locations can be extended continuously across an image. Interpolating the PV-correction coefficients A and B is convenient[4]; however, the dose estimate depends directly on the corrected PV rather than on the fitted coefficients. We therefore hypothesized that directly interpolating the PV correction would provide a more direct representation of the LRA.

This study had four objectives: (1) to characterize the LRA of EBT4 and EBT-XD films; (2) to compare the Lewis-type coefficient-interpolation LRA correction[4] using two films (Lewis LRA correction) with a film-cyclic correction based on direct interpolation in PV (Full LRA correction); (3) to quantify the performance of single-, double-, and triple-channel dosimetry under the No LRA, Lewis LRA, and Full LRA correction conditions; and (4) to evaluate central-dose agreement after applying the Lewis and Full corrections. This comparison was intended to inform the implementation and assessment of LRA correction in radiochromic film dosimetry.

## 2. MATERIALS AND METHODS

### 2.1 Film, scanner, and image acquisition

Gafchromic EBT4 (Lot 04082502) and EBT-XD (Lot 01292404) films (Ashland, Bridgewater, NJ, USA) were evaluated. Films were digitized in transmission mode with an Epson Expression 12000XL-PH flatbed scanner (Seiko Epson Corp., Nagano, Japan). Images were acquired as 48-bit RGB TIFF files at 150 dpi with all scanner color and image corrections disabled. The scanner was warmed with repeated prescans before data acquisition. Films were kept flat at reproducible locations on the scan window, and the original sheet orientation was marked before cutting and preserved throughout irradiation and readout. Images of both EBT4 and EBT-XD for both calibration and measured sets were acquired approximately 22 h after irradiation. Subsequent image processing, calibration, LRA fitting, dose reconstruction, profile and dose analyses, and statistical testing were performed in MATLAB R2024b (MathWorks, Natick, MA).

### 2.2 Measurement of the lateral response artifact

Seven 5 × 5 $cm^2$ film pieces irradiated with different dose levels were used to characterize the LRA. The film pieces were arranged at seven discrete lateral positions in the *y* direction of the scanner and shifted cyclically over seven scans so that each film was measured once at every position (Figure 1). The entire lateral region of the scanner was covered, and seven positions were chosen as a trade-off between accuracy and efficiency. Position 4 was centered on the scanner midline along the LRA axis and served as the reference. This crossed design separates film-to-film variation from scanner-position dependence because each off-center measurement can be paired with the center measurement of the same film piece.

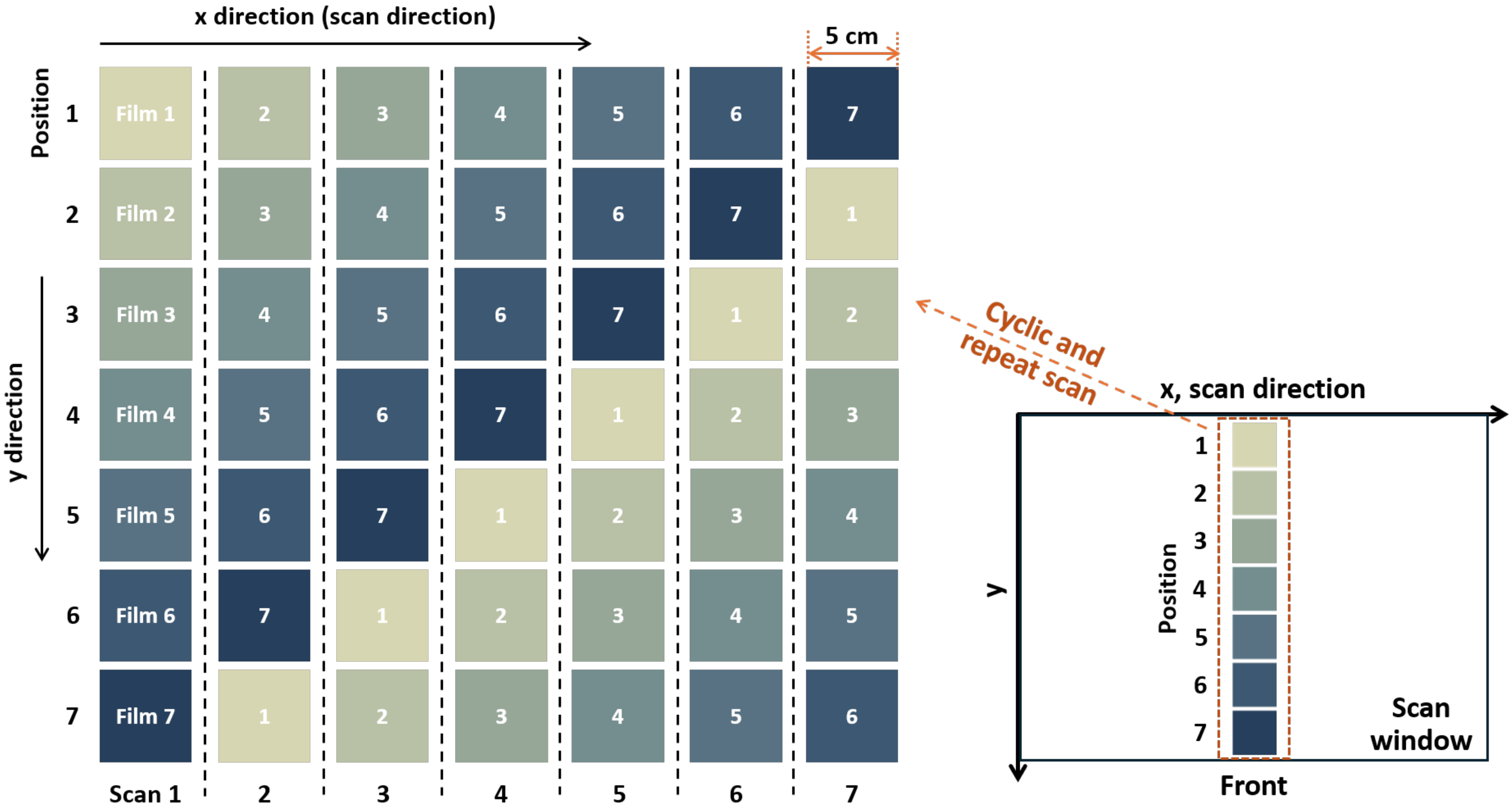

**Figure 1.** Cyclic-shift acquisition to quantify the lateral response artifact. Seven film pieces irradiated at different dose levels were scanned at seven lateral positions over seven scans, such that each film piece was measured once at every position. Position 4 was centered on the scanner midline and served as the reference.

For each film and color channel, the mean PV and standard deviation were calculated within a central 5 × 5 $mm^2$ region of interest (ROI) at each position. Seven ROIs corresponding to each scan position were fixed relative to the scanner, and the film center was aligned with each ROI in turn. At each off-center position $i$ and channel $k$, the center response $PV_{c,k}$ was related to the local response $PV_{i,k}$ using a Lewis-type affine model[4]:

$$PV_{c,k} = A_{i,k} + B_{i,k} \cdot PV_{i,k} \tag{1}$$

where $A_{i,k}$ and $B_{i,k}$ are position- and channel-specific coefficients. The additive correction that transforms the local response to the center-equivalent response is therefore:

$$PV_{c,k} - PV_{i,k} = A_{i,k} + (B_{i,k} - 1) \cdot PV_{i,k} \tag{2}$$

At the scanner center, the identity constraints $A_{4,k} = 0$ and $B_{4,k} = 1$ were imposed. Linear fits were performed independently for the red, green, and blue channels of EBT4 and EBT-XD. Residual patterns, coefficient trends, and residual symmetry (Sec. 3.1) were examined in PV space.

**2.3 Position-continuous LRA correction in PV space**

The discrete correction in Eq. (2) must be evaluated at every image pixel, including locations between the seven sampled scanner positions. Three conditions were evaluated: no LRA correction (No LRA), Lewis LRA correction (Lewis LRA), and Full LRA correction (Full LRA). Under no correction, the measured RGB image was passed directly to dose reconstruction without any PV correction.

For the Lewis LRA correction[4], two films, including a nonirradiated film for simplicity, were used to determine $A_{i,k}$ and $B_{i,k}$ in Eq. (1) at each sampled position and channel. The resulting $A_{i,k}$ and $B_{i,k}$ values were then linearly interpolated along the scanner's lateral direction $y$ to obtain the spatially continuous PV correction using Eq. (2).

For the Full LRA correction, all valid center-local PV pairs were used to estimate $A_{i,k}$ and $B_{i,k}$ at each off-center position and channel. Unlike the Lewis correction, which linearly interpolates $A_{i,k}$ and $B_{i,k}$, the Full LRA correction directly interpolates the additive PV correction applied before dose reconstruction. For each channel and sampled position, the PV correction $C_{Full,i,k}$ was first evaluated using Eq. (2) at three fixed PV anchors ($PV_j$ = 10,000, 30,000, and 50,000) spanning the measured response range:

$$C_{Full,i,k}(PV_j) = A_{i,k} + (B_{i,k} - 1) \cdot PV_j, \qquad PV_j = 1, 3, 5 \times 10^4 \tag{3}$$

Position 4 was assigned $C_{Full,4,k} = 0$. For each channel and PV anchor, $C_{Full,i,k}$ was then interpolated along $y$ using shape-preserving piecewise cubic Hermite interpolation (PCHIP). At a given lateral location, the correction for intermediate PVs was linearly interpolated between the three anchors, consistent with the linear dependence of the LRA correction on PV[4]. Corrections were restricted to the experimentally measured PV range to avoid uncontrolled extrapolation.

After those corrections were applied, the center-equivalent RGB image was passed to the same single-, double-, or triple-channel dose-reconstruction algorithm used for the uncorrected image. This ordering separates correction of the scanner response from subsequent single-, double-, or triple-channel film dose reconstruction.

**2.4 Film calibration**

The scanner response varies along the lateral-response axis ($y$) but not along the lamp-motion axis ($x$) (Supplementary Figure S1). To isolate the effect of the LRA on film profiles, each film was scanned in both the landscape and portrait orientations (Fig. S1). Using the same film for both scans minimized confounding effects from film nonuniformity, dust, fingerprints, and other film-specific artifacts. Because film orientation is known to affect scanner response[1], separate calibration curves were established for the landscape and portrait orientations.

Calibration films were irradiated with an 18 MeV electron beam using a 10 × 10 cm$^2$ cone on a Varian 2100EX linear accelerator[18]. Films were positioned at 2 cm depth in solid water with 8 cm of backscatter material. All calibration films were scanned at Position 4, on the midline of the scanner's lateral-response axis. Separate calibration curves were generated for landscape and portrait scans of each film model using the same calibration films, scanner settings, and postirradiation interval for the corresponding profile measurements (Fig. S2).

Mean RGB PV and standard deviation were extracted in the central 5 × 5 mm$^2$ of each calibration film. The calibration data were fitted with the following function[7]:

$$D = \frac{p_1 - p_2 \cdot PV}{PV - p_3} \quad (4)$$

where $p_1$, $p_2$, and $p_3$ are fitting parameters. Uncertainties in both dose and PV were incorporated in the calibration fit using the effective-variance method[17].

**2.5 Dose-reconstruction methods**

Seven dose-reconstruction methods were evaluated under each PV treatment: single-channel red (R), green (G), and blue (B); double-channel red-blue (RB) and green-blue (GB)[6]; triple-channel Micke (RGB Micke)[6] and the Pérez Azorín triple-channel method without film prescan (RGB Azorín)[7]. The same orientation-specific calibration data were used for all reconstruction methods within a given scan orientation.

**2.6 Profile measurement and analysis**

Film pieces measuring 2.5 × 25 cm$^2$ were irradiated with the same 18 MeV electron beam using a 20 × 20 cm$^2$ cone to measure the beam profiles. The same measurement depth and backscatter material as the calibration films were used. EBT4 measurements were acquired at

100, 500, and 1000 MU, while EBT-XD measurements were acquired at 100, 500, 1000, 2000, and 3000 MU, corresponding to their recommend dose range.

At each MU level, the same irradiated film was scanned in landscape and portrait orientations (Fig. S1), and the corresponding orientation-specific calibration curve was applied. Identical preprocessing was used for all methods and correction conditions. Isolated extreme values were identified using robust percentile limits; local spikes were detected relative to a 31-point moving median; and the remaining profile was smoothed with a 15-point moving median. Each profile was independently normalized to its mean reconstructed dose within a 5-mm-wide region centered on the central axis.

The signed relative difference between the portrait and landscape profiles, together with the mean absolute relative difference, was used to evaluate the performance of the three correction conditions in combination with the seven film-dosimetry methods:

$$\delta(y) = [P_{Por}(y) - P_{Lan}(y)]/P_{Lan}(y) \times 100\% \tag{5}$$

$$E = mean\{|\delta(y)|, |y| \leq 100\ mm\} \tag{6}$$

The interval $|y| \leq 100$ mm corresponds to the nominal width of the 20 × 20 $cm^2$ field.

## 2.7 Central-dose agreement analysis

In addition to profile consistency, central-dose agreement was evaluated for the portrait scans, for which the long film dimension was aligned with the scanner lateral-response axis *y* (Fig. S1). For each film model, MU level, dose reconstruction method, and LRA correction approach, the mean reconstructed dose was calculated within a 5 × 5 $mm^2$ ROI centered on the aligned scanner and film centers. The reference dose at the measurement depth was calculated from the LINAC output, percentage depth dose, solid-water-to-water correction, and electron-cone output factor[19]. Because each condition was represented by one film measurement, this analysis was intended to compare dose-agreement trends rather than to establish absolute dosimetric accuracy or repeatability.

## 2.8 Declaration of Generative AI and AI-assisted technologies in the writing process

During the preparation of this work the authors used OpenAI Codex (OpenAI, San Francisco, California, USA) to assist with editing and language polishing. All content generated using this tool was subsequently reviewed and revised by the authors, who take full responsibility for the final content of the publication.

# 3. RESULTS

## 3.1 PV-domain characterization of LRA

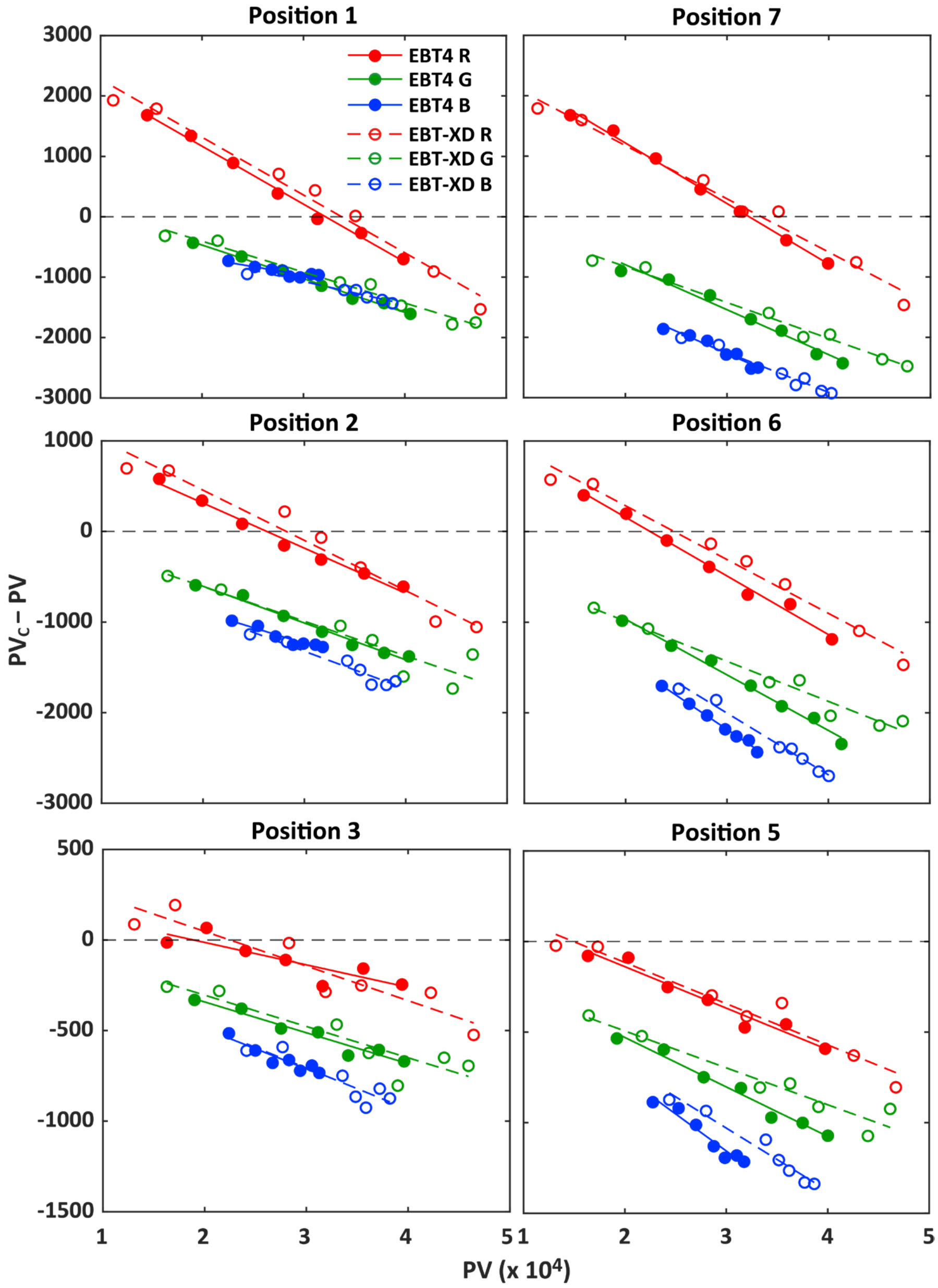

**Figure 2.** PV-domain lateral response artifact for EBT4 and EBT-XD films. The additive center-equivalent correction, $PV_c - PV$, is plotted against local PV for EBT4 and EBT-XD in the red, green, and blue channels. Panels correspond to noncentral scanner positions; position 4 was the reference center and is not shown.

For both EBT4 and EBT-XD films, $PV_c - PV$ was approximately linear with local PV at every off-center position and in all three channels (Figure 2). The sign and magnitude of the correction varied with lateral position and PV. For the blue and green channels, $PV_c - PV$ remained negative across the measured positions and PV ranges, whereas the red-channel correction could be positive or negative depending on PV and position. The largest corrections for each channel generally occurred at the outer positions, whereas positions adjacent to the center required smaller adjustments.

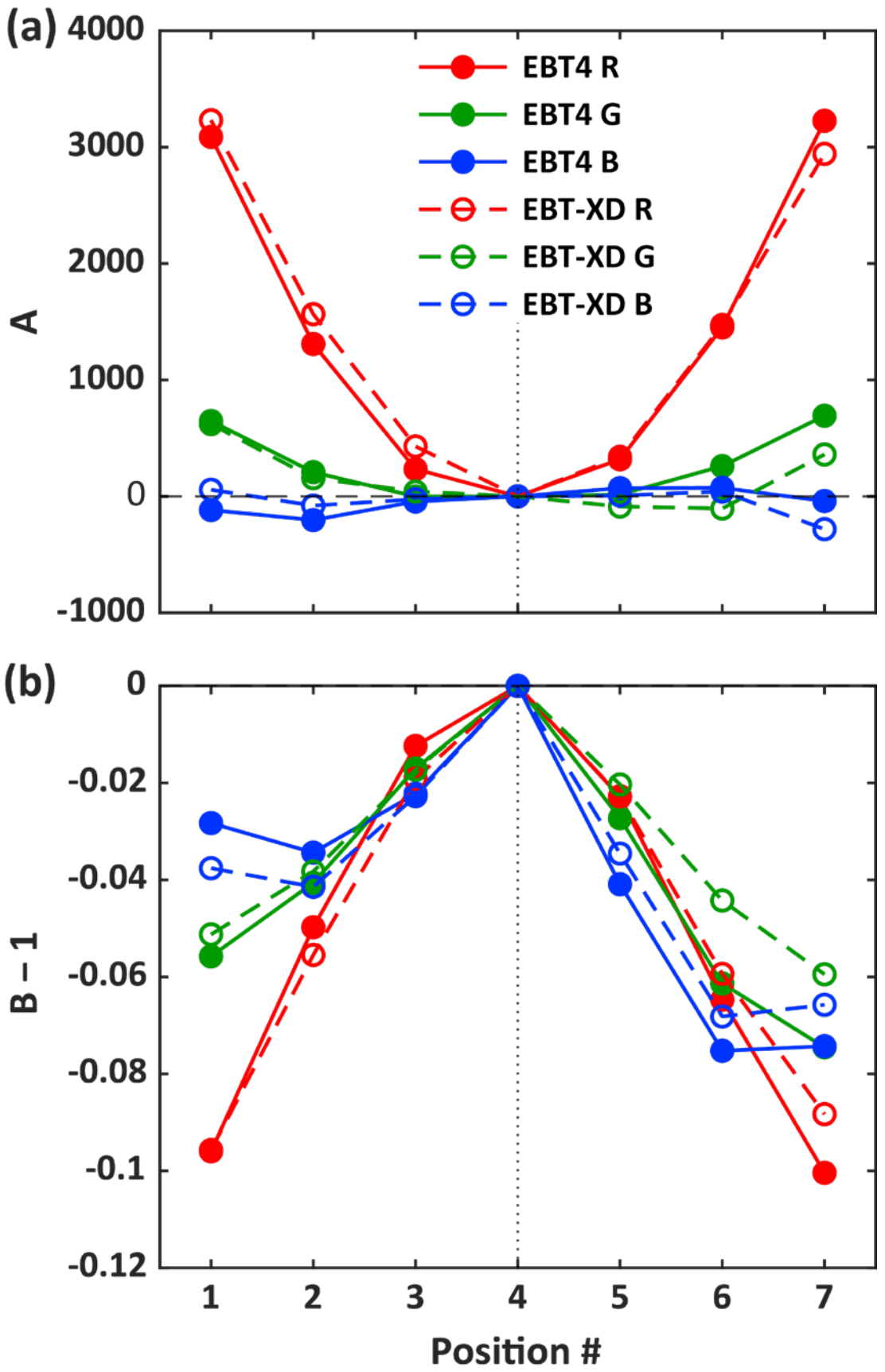

**Figure 3.** Position dependence of the affine LRA correction coefficients for EBT4 and EBT-XD films: (a) intercept A and (b) slope offset B − 1. Position 4 was constrained to A = 0 and B − 1 = 0.

The LRA was asymmetric about the scanner center. Fitted A and B − 1 values differed between opposite sides of the scanner, with the most pronounced side-to-side differences occurring in portions of the blue-channel response (Fig. 2-3).

The exploratory nested linear-model F-tests indicated film-model dependence in the LRA relationship (Supplementary Section 2 and Fig. S3). Because the differences were not described by a consistent offset between EBT4 and EBT-XD, these data support using separate LRA correction models for the two film types in this workflow.

We next examined how the LRA PV correction varied continuously with scanner position. The Full and Lewis LRA corrections exhibited similar overall spatial trends but differed both at the sampled calibration positions and between them (Fig. 4). At the calibration positions, these differences arose because the Full LRA correction method used all available PV pairs, whereas the Lewis LRA correction relied on only two films. Between calibration positions, additional differences resulted from directly applying PCHIP interpolation to C in the Full LRA correction method, rather than linearly interpolating A and B − 1. Consequently, the Lewis

LRA correction method imposed an approximately linear variation in the PV correction between calibration positions, whereas the physical spatial dependence of the LRA may be nonlinear and exhibit curvature with distance from the scanner center[3]. The largest discrepancies were observed near the outer scanner positions and were generally most pronounced in the blue channel.

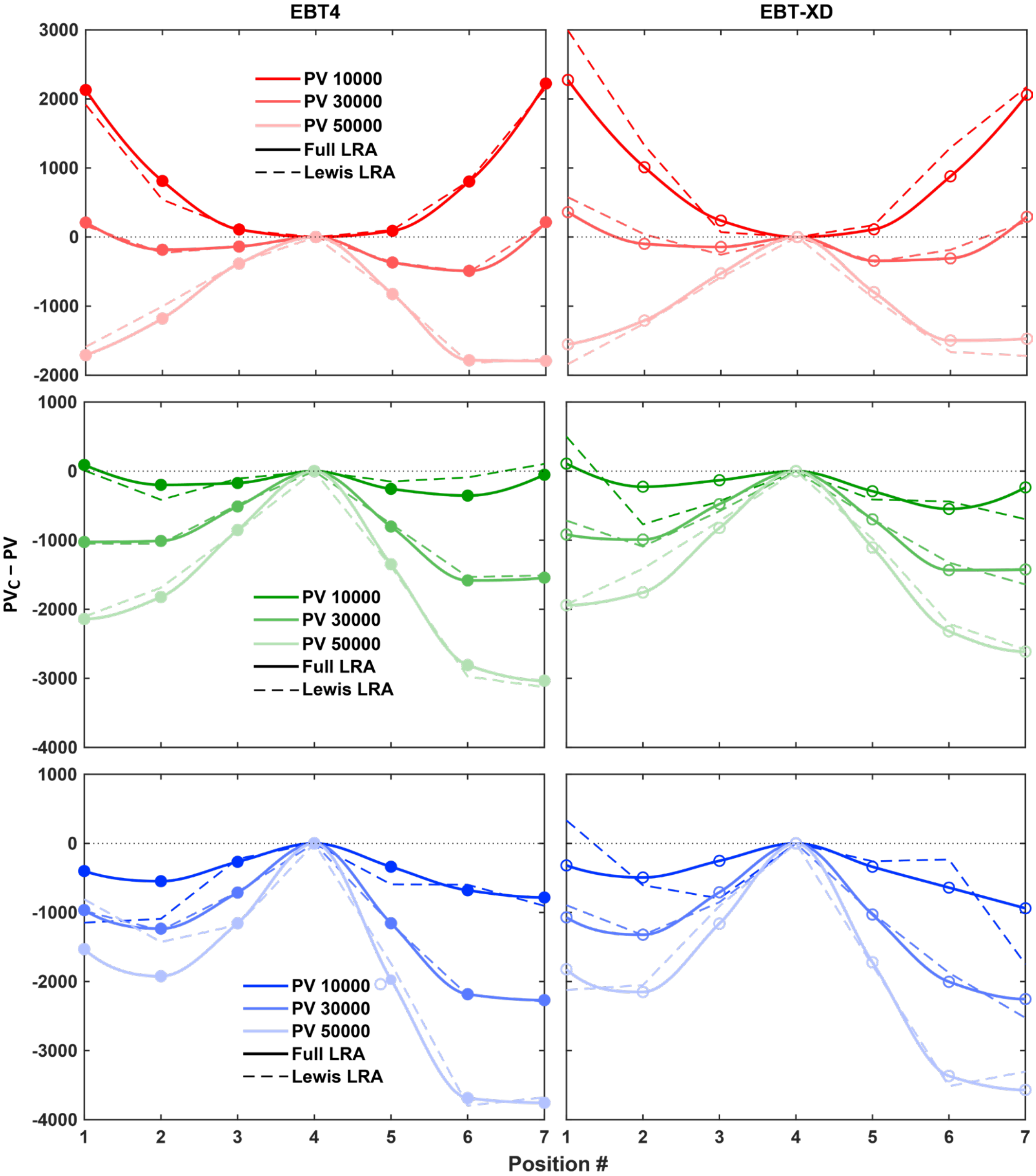

**Figure 4.** Comparison of the Full PCHIP LRA correction and the two-film Lewis LRA correction for EBT4 and EBT-XD films. Rows show the red, green, and blue channels; columns show the two film models. Curves are displayed at PV anchors of 10,000, 30,000, and 50,000. Solid curves use all available center-local pairs to fit A and B and apply PCHIP directly to $C(y,PV) = PV_c - PV$. Dashed curves use only two films to determine A and B and then linearly interpolate A and $B - 1$. Position 4 was constrained to $C = 0$.

### 3.2 Film calibration performance

All orientation- and channel-specific calibration curves showed high goodness of fit (Supplementary Fig. S2 and Table S1), with $R^2$ values from 0.9982 to 0.9999. The lowest $R^2$ values occurred in the blue channel for both film models. Landscape and portrait calibration curves had similar overall shapes but distinct fitted parameters.

The EBT4 calibration range extended to approximately 1000 cGy, whereas the EBT-XD range extended to approximately 4000 cGy. The red channel provided the greatest response sensitivity at low-to-moderate doses. The blue channel retained higher PV at large doses but showed comparatively low sensitivity over portions of the evaluated range (Fig. S2).

### 3.3 Profile consistency for 1000 MU EBT4

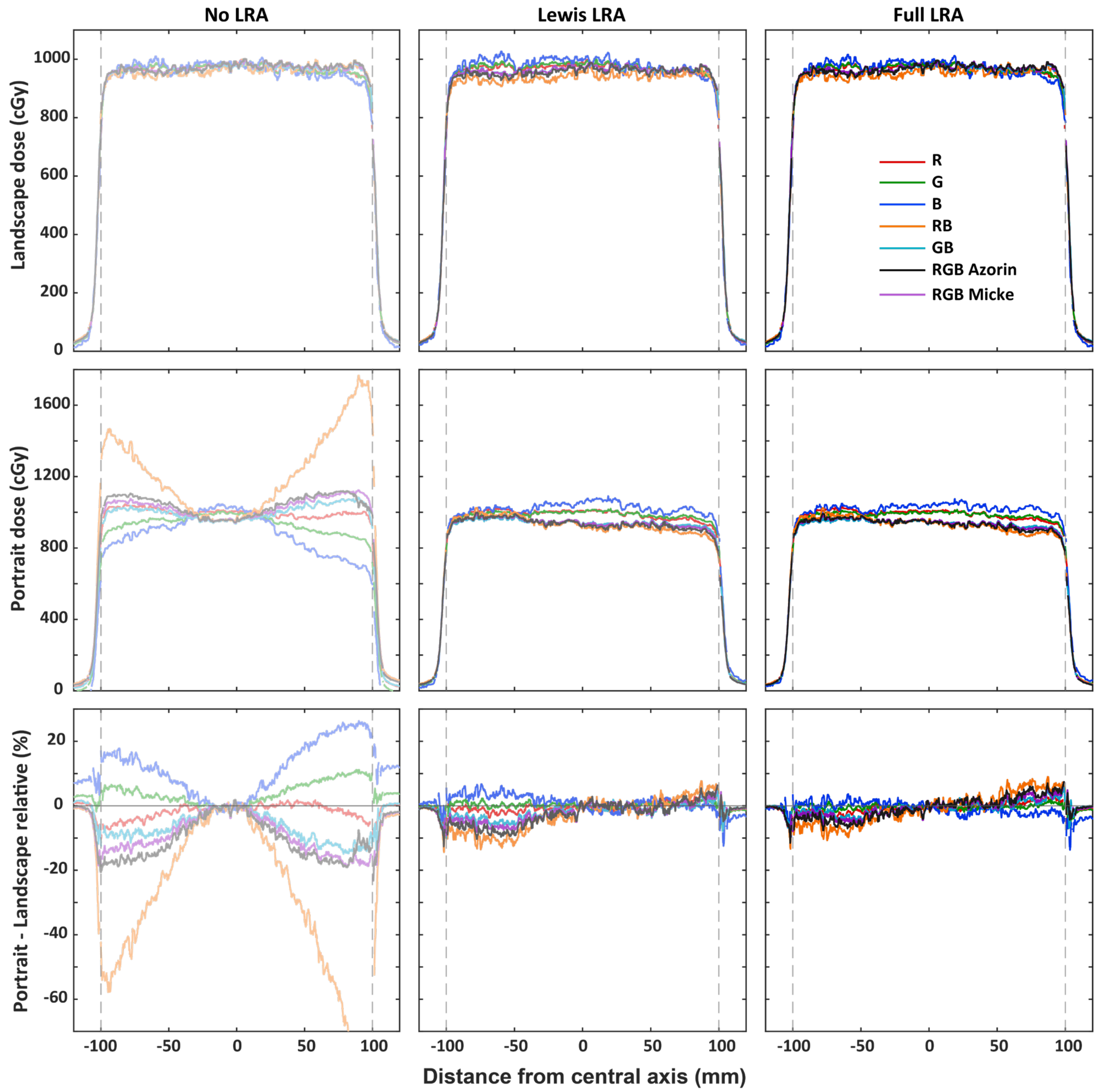

**Figure 5.** Dose profiles reconstructed from an EBT4 film irradiated with 1000 MU using an 18 MeV, $20 \times 20$ cm$^2$ electron field. Columns show No LRA, Lewis LRA, and Full LRA corrections. Rows show landscape absolute dose, portrait absolute dose, and the portrait-landscape relative profile difference $\delta(y)$.

For the representative 1000-MU EBT4 irradiation, uncorrected landscape profiles were comparatively uniform across the irradiation field, whereas portrait profiles showed method- and channel-dependent curvature caused by LRA across the field (Figure 5, left column). For the single-channel methods, the R portrait profile was closest to the landscape profile, whereas G and especially B showed larger off-axis differences. For the double- and triple-channel methods, RB and both triple-channel methods retained substantial lateral position dependence. Both explicit corrections brought the portrait profiles substantially closer to the landscape profiles (Fig. 5, middle and right columns).

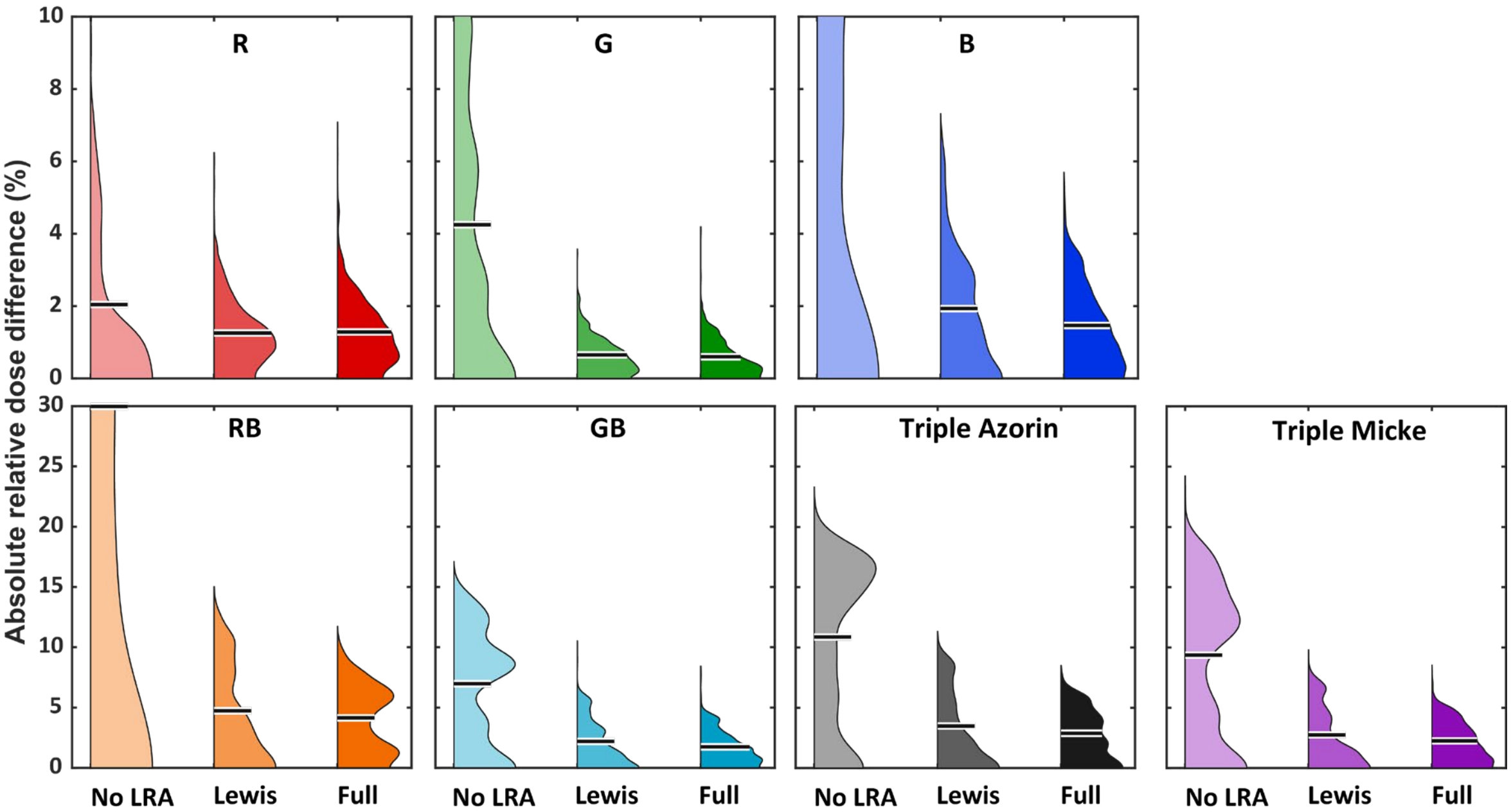

**Figure 6.** Half violin plots of the absolute portrait-landscape relative profile difference $|\delta(y)|$ within the nominal $20 \times 20$ cm$^2$ field ($|y| \leq 100$ mm) for the 1000-MU EBT4 irradiation shown in Fig. 5. Each panel compares No LRA, Lewis LRA, and Full LRA corrections for one reconstruction method. Horizontal bars indicate the median of each distribution.

Figure 6 presents half-violin plots of the pixelwise absolute portrait-landscape relative profile difference $|\delta(y)|$ ($\delta(y)$ defined in Eq. 5) for the three correction conditions across the seven film-dosimetry methods. The distributions show reductions in both the median and range of $|\delta(y)|$ after Lewis LRA and Full LRA corrections. Before correction, R showed the smallest discrepancy among the single-channel methods. Among the double-channel methods, GB

showed lower discrepancies than RB. Neither triple-channel method consistently outperformed R or GB. After both corrections, G showed smaller profile discrepancies than R and B. GB and the two triple-channel methods all had mean $|\delta(y)|$ values of 2%-3% for the 1000-MU EBT4 measurement.

**3.4 Method-level profile consistency**

For each film model and dose-reconstruction method, method-level portrait-landscape profile consistency was summarized as the arithmetic mean of $E$ (Eq. 6) across the evaluated MU levels. This comparison assessed the consistency between the portrait profile after LRA correction and the landscape profile, and was not an independent validation of absolute-dose accuracy.

For EBT4, the uncorrected portrait-landscape discrepancy was strongly method dependent (Table 1). R showed the greatest inherent consistency, whereas B and RB showed the largest discrepancies. Both Lewis and Full LRA corrections substantially improved profile consistency, particularly for G, B, and RB. After correction, mean $E$ was below 4% for all methods except B. The two corrections produced nearly identical EBT4 results, differing by no more than 0.3% in mean $E$ for any method, and neither was consistently superior.

**Table 1**. Mean absolute portrait-landscape relative profile difference, $E$ (%), within the nominal 20 × 20 cm² field ($|y| \leq 100$ mm), averaged across the evaluated MU levels.

| Method | Mean of $E$ (%) | | | | | |
|---|---|---|---|---|---|---|
| | EBT4 | | | EBT-XD | | |
| | No LRA | Lewis LRA | Full LRA | No LRA | Lewis LRA | Full LRA |
| R | 3.1 | 1.5 | 1.8 | 3.0 | 2.9 | 3.0 |
| G | 10.4 | 1.4 | 1.4 | 14.6 | 3.0 | 2.4 |
| B | 28.6 | 5.1 | 5.3 | 44.4 | 6.7 | 6.1 |
| RB | 19.0 | 3.6 | 3.5 | 34.2 | 6.8 | 4.5 |
| GB | 4.9 | 2.1 | 2.1 | 6.5 | 3.1 | 2.9 |
| RGB Azorín | 9.4 | 2.8 | 2.7 | 14.6 | 3.9 | 3.0 |
| RGB Micke | 9.6 | 2.5 | 2.4 | 14.2 | 4.8 | 3.4 |

EBT4 averages include 100, 500, and 1000 MU; EBT-XD averages include 100, 500, 1000, 2000, and 3000 MU. Per-MU $E$ values for EBT4 and EBT-XD are listed in Supplementary Tables S2 and S3.

EBT-XD showed the same overall method dependence but larger uncorrected discrepancies than EBT4, particularly for G, B, RB, and the triple-channel methods. R again showed the greatest consistency, while B had the largest discrepancy. Both corrections markedly reduced orientation dependence for all methods except R, whose uncorrected discrepancy was already small. Full LRA generally produced lower residual discrepancies than Lewis LRA, with the clearest additional improvement for RB, RGB Micke, and RGB Azorín. B retained the largest residual discrepancy after correction. Across the combined-channel methods, GB consistently showed lower discrepancies than RB. The triple-channel methods improved substantially after explicit correction but did not consistently outperform R or GB.

Overall, explicit LRA correction substantially improved portrait-landscape profile consistency, especially for methods involving the green and blue channels. Lewis and Full LRA correction performed comparably for EBT4, whereas Full correction produced lower residual discrepancies for several double- and triple-channel methods with EBT-XD.

**3.5 Central-dose agreement under the three correction conditions**

Figure 7 shows the difference between the portrait-film center dose and the reference LINAC output. Because each condition was represented by one film measurement, the comparison was used to evaluate dose-agreement trends among reconstruction methods and LRA corrections rather than to establish absolute dosimetric accuracy or repeatability.

The largest relative dose deviations occurred at 100 MU, particularly for B and the blue-influenced double- and triple-channel methods. At this MU level, both Lewis and Full LRA corrections increased the deviation of several single-channel estimates relative to No LRA correction. Compared with Lewis LRA correction, Full LRA correction generally produced smaller dose deviations, most notably for B in both film models.

At 500 MU and above, R showed the most consistent agreement with the reference output and little difference between Lewis and Full LRA corrections. G generally retained a positive dose bias after correction, whereas B showed even larger dose deviations. Among the double-channel methods, GB showed smaller absolute dose deviations than RB for both film models and all three correction conditions. RB tended toward a negative dose bias at higher MU, particularly with Lewis LRA correction, while Full LRA correction generally reduced the magnitude of this dose bias.

The triple-channel methods varied less than B and RB at moderate-to-high MU. For EBT4, Lewis and Full LRA corrections produced broadly comparable results. For EBT-XD, Full LRA correction generally improved agreement relative to Lewis LRA correction for RB and RGB Micke, whereas the difference for RGB Azorín was smaller and MU dependent. Overall, R, GB, and the two triple-channel methods showed the most favorable dose agreement trends.

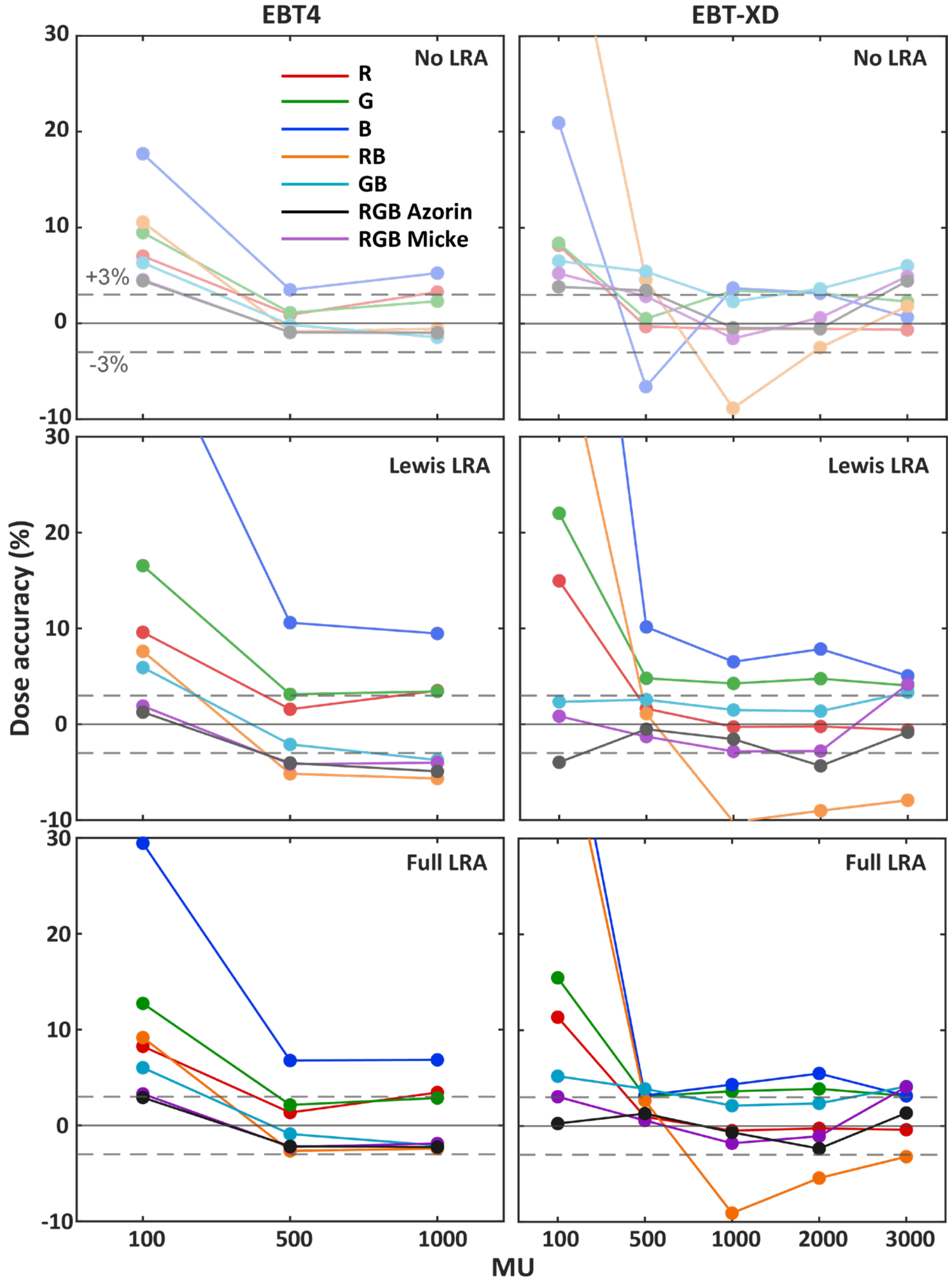

**Figure 7**. Relative dose difference between the portrait-film center dose and the LINAC reference output for EBT4 and EBT-XD. The mean dose within a 5 × 5 $mm^2$ ROI centered on the radiation field was evaluated. Dashed lines indicate ±3%. Each point represents one film measurement; error bars are therefore not shown. Some low-MU relative-difference values fall outside the displayed y-axis range.

## 4. DISCUSSION

The principal finding of this study is that explicit correction of the raw RGB response before single- or multichannel dose reconstruction improved portrait-landscape profile consistency. This ordering separates two different sources of variation: the position- and channel-dependent lateral response artifact of the scanner-film system and the channel-common disturbances minimized by multichannel dose-reconstruction models. We evaluated the effects of the Lewis and Full LRA corrections on lateral profile consistency and central-dose agreement across single-, double-, and triple-channel film dosimetry methods.

The affine relationship between center and local PV (Fig. 2) agrees with the framework of Lewis and Chan[4]. The present study extends that framework by introducing a full-data correction that directly interpolates the PV correction, comparing EBT4 with EBT-XD, and evaluating both correction approaches across seven single-, double-, and triple-channel methods. The cyclic-shift design also provides within-film pairing between each off-center response and the corresponding center response, thereby reducing confounding from film-to-film response variation (Fig. 1).

Previous work has shown that explicit scanner correction can recover large-field film profiles and that restricting readout to central scanner regions through image stitching can reduce LRA[5,12]. Accelerated characterization protocols further demonstrate that irradiation nonuniformity can be incorporated into an apparent scanner artifact if it is not independently accounted for[13]. The present acquisition was designed to reduce irradiation-related bias in the LRA estimate by cycling the same set of films through all sampled positions. The correction was both film-model specific (Fig. S3) and asymmetric about the scanner center (Fig. 2-3). EBT4 and EBT-XD differ in active-layer response, spectral characteristics, dose sensitivity, and orientation dependence[11,14-16], so a scanner-induced LRA may not produce the same PV artifact and subsequent dose bias. In addition, the observed asymmetry in LRA argues against describing it with a symmetric parabola function[3]. The nested-model analysis (Sec. 2 of Supplementary Material) does not imply that a difference was detected at every position-channel pair, but it indicates that these data do not support a shared EBT4/EBT-XD correction. Commissioning should therefore be specific to the film model and readout protocol and, when tighter uncertainty requirements apply, to the production lot.

The comparison of Lewis and Full LRA corrections illustrates the tradeoff between calibration burden and response-range coverage. The Lewis-type approach requires two films and performed essentially as well as Full LRA for EBT4 (Table 1). Its simplicity is attractive for routine commissioning, but the fitted relationship is determined entirely by two endpoint measurements and is therefore sensitive to endpoint noise or poor representation of the intervening PV range. Full LRA correction uses more extensive PV-response data and interpolates the PV correction applied to the image rather than separately interpolating its fitted coefficients. It produced lower observed discrepancies for several EBT-XD double- and triple-channel methods, for which small residual channel differences had a larger effect on the reconstructed dose.

The relative performance of the dose-reconstruction methods after LRA correction depended more on channel compatibility than on the number of channels (Fig. 5-6, Table 1). R showed the smallest uncorrected discrepancy among the single-channel methods. Among the double-channel methods, GB consistently showed lower discrepancies than RB, indicating that combining channels is beneficial only when their LRA residuals are sufficiently compatible with the assumptions of the solver. The triple-channel methods improved markedly after explicit LRA correction but did not consistently surpass R or GB. Thus, including more channels does not guarantee improved performance when the scanner artifact is channel dependent.

The blue channel was the main source of residual instability. Its absolute PV correction was not always the largest, yet B produced the largest profile discrepancies and the most pronounced low-MU center-dose deviations. This apparent mismatch can be understood through the calibration curve (Fig. S2): in a low-sensitivity region, a modest residual PV error may produce a large dose change. Blue-channel information can extend dynamic range and support estimation of film-thickness-related disturbances, but its contribution should be weighted against the local calibration sensitivity and differences among residual LRA effects in the RGB channels. In the present data, blue-channel behavior limited RB and prevented the triple-channel solutions from consistently improving on R.

Portrait-landscape profile agreement and central-dose agreement provide complementary, not interchangeable, tests. The scan orientation comparison used the same irradiated film and field under two scan orientations, providing a within-film assessment of profile agreement (Fig. S1). Inclusion of the entire nominal field width, including the penumbrae, also made the metric responsive to edge distortion (Fig. 5). By contrast, the center-dose analysis (Fig. 7) evaluated the combined effect of PV correction, film calibration, and the channel-coupled dose solver at one location. A correction could therefore improve profile shape while shifting the reconstructed center dose (Figs. 5 and 7).

This distinction was most evident at low MU. Lewis and Full LRA corrections occasionally increased the center-dose deviation of single-channel estimates, particularly B, even though both corrections improved profile consistency (Fig. 7). For RB, correction also produced a negative bias at higher MU, with Full LRA correction generally reducing the magnitude relative to Lewis. These observations do not constitute an absolute-accuracy comparison because each condition was represented by one film and the reference output was calculated rather than measured concurrently with an independent detector. They nevertheless illustrate that a spatially successful correction can interact with a nonlinear dose conversion or multichannel solver in a method- and dose-dependent manner.

Several limitations of this work should be considered. Only one scanner and one production lot of each film model were evaluated, and the results may not transfer directly to other scanners and film lots. The central-dose comparison used one film per condition and did not quantify measurement uncertainty or repeatability.

For practical implementation, the results support a modular commissioning workflow. Orientation-matched calibration should first be established, followed by LRA characterization for the specific scanner, film model, and RGB channel. The Lewis two-film LRA correction may be selected when its observed profile and dose agreement meet the intended tolerances and its calibration films span the working PV range. The Full LRA correction may be preferable when lower residual profile or dose discrepancies are required or when combined-channel EBT-XD reconstruction is used. Each raw channel should then be transformed to a center-equivalent PV image before dose reconstruction. Finally, both spatial profile consistency and dose agreement should be verified over the intended dose and field-size ranges.

## 5. CONCLUSIONS

Flatbed-scanner LRA was approximately affine in PV at each sampled position but varied with scanner side, lateral position, color channel, PV, and film model. Explicit correction of the raw RGB response substantially improved profile consistency between scan orientations with and without LRA, whereas multichannel reconstruction alone did not eliminate the artifact. The two-film Lewis-type LRA correction performed comparably to Full LRA correction for the EBT4 conditions studied, whereas the full-data correction produced lower residual discrepancies for several EBT-XD double- and triple-channel methods. Residual blue-channel behavior remained the principal limitation, and improved spatial consistency did not necessarily produce improved center-dose agreement. LRA correction should therefore be commissioned as a film- and channel-specific preprocessing step, with profile consistency and central-dose agreement verified independently.

## DATA AVAILABILITY STATEMENT

Data are available from the corresponding author upon reasonable request.